# Rejoinder: Citation Statistics

**Robert Adler, John Ewing and Peter Taylor**

We would like to thank the discussants for reading our report and for their insightful and constructive comments.

To start our brief response, we would like to quote Bernard Silverman's phrase "reducing an assessment of an individual to a single number is both morally and professionally repugnant." Bernard puts it strongly, but his underlying point, with which we strongly agree, is that "research quality" is not something that ought to be regarded as well-ordered.

We note the general support for the case that any analysis should be carried out in the context of a properly-defined model. Peter Hall calls for statisticians to undertake a study of "the nature of citation data, the information they contain and methods for analysing them if one must." Among the three of us, there are varying levels of enthusiasm for advocating such a project. A possible downside is the danger that such a study will add to the burgeoning number of proposals for carrying out citation analysis in a "better" way, and none of us have much enthusiasm for this. On the plus side, such a study would enable the mathematical sciences community to comment more authoritatively on citation statistics and the quantitative ranking measures that are derived from them. Given that the scientometric industry shows every sign of growing, it can be argued that it is the responsibility of the mathematical sciences, and particularly of statisticians, to develop this capability.

David Spiegelhalter and Harvey Goldstein pointed out that there is a lack of independence between individual authors' citation records due to issues of co-authorship. The effects of this lack of independence seem to be very poorly understood, and nothing in the literature that we reviewed sheds any light on them.

In our report, we spent some time discussing the meaning of citations. Sune Lehmann, Benny Lautrup and Andrew Jackson took this point further in their discussion of the fact that there needs to be agreement on the basic meaning of a researcher's citation distribution, which is something that goes beyond merely knowing what citations mean, which itself is not clear. Their example involving researchers A and B makes this point clearly.

We would like to emphasise three final points that have more to do with human behavior than statistics, and which were not emphasised in the report itself. The first is related to Bernard Silverman's point that any measurement or ranking system will drive researcher behavior via natural feedback mechanisms. Traditionally, the mechanisms adopted in academia have been qualitative rather than quantitative. Peer review has been at the core of the system. When carefully done, peer review not only provides accurate and professional assessments of an individual's contributions, but it also provides a balanced and educated interpretation of quantitative information such as prizes and citation data. Moving to a system based purely on quantitative citation metrics will deliver feedback more frequently, more unequivocally, and in a different way. It is not at all clear that "good research" (and we realise how loaded this term is) will be encouraged by such a system. Our strong opinion is that this feedback aspect is very important.

Related to this issue is another of particular concern. In general, it is not all that easy to fool one's peers, but it takes little imagination to see how, by adopting citation policies that are different from the norm in a particular discipline or sub-discipline, a small group of individuals could easily fool an automated assessment system built on citation data. Assessment is important to all of us, as individuals, as institutions, and as representatives of disciplines. Adopting a system, for short term gains, that is so easily open to abuse is a risk to research standards in the long term.

*Robert Adler, Faculty of Electrical Engineering, Faculty of Industrial Engineering and Management, Technion, Haifa, Israel, 32000 e-mail:*
*robert@ieadler.technion.ac.il. John Ewing, President, Math for America 800 Third Ave, 31st fl, New York, New York 10022, USA e-mail:*
*ewing@mathforamerica.org. Peter Taylor, Department of Mathematics and Statistics, University of Melbourne, Vic 3010, Australia e-mail: p.taylor@ms.unimelb.edu.au.*

Our final point, which has been amplified by our experiences since the report was first released, is that almost everyone is affected by conflicts of interest when the topic of research assessment comes up. For most of us, the way our research is regarded goes to the very core of our professional identity, and it would be a rare individual who could isolate his or her opinion about a particular method of research assessment and the way that his or her own research is ranked by the method. For example, most people who do well according to h-indices tend to think that the h-index is not a bad measure of researcher quality. There are also individuals who have built careers, and companies that have profited, from undertaking research assessment in a particular way. Since we are certain that it is healthy for all disciplines to have a multitude of skills and temperaments in their research communities, this observation leads us back to where we started: "research quality" is an inherently multidimensional object and should be treated as such.